\documentclass[pdf,12pt]{amsart}
\usepackage{times}
\usepackage{curves}
\usepackage{amssymb,amsfonts,eufrak,amsmath,amscd}
\usepackage[pdftex]{graphicx}
\begin{document}

\newtheorem{theo}{Theorem}
\newtheorem{lem}{Lemma}
\newtheorem{cor}{Corollary}
\newtheorem{prop}{Proposition}
\newtheorem{ex}{Example}
\newtheorem{remar}{Remark}
\newtheorem{rul}{Rule}
\newtheorem{conj}{Conjecture}
\newtheorem{defi}{Definition}

\newcommand{\bt}{\begin{theorem}\em}
\newcommand{\et}{\end{theorem}}
\newcommand{\petet}[2]{{}^{#2}\!{#1}}
\newcommand{\side}[2]{{}^{#1}\!{#2}}

\newcommand{\mop}[2]{\stackrel{{}_{{#2}}}{#1}}
\newcommand{\Mop}[4]{\overset{\ \ \ #4}
{\underset{\ \ \ #3}
{\mathop{\sideset{^{{}_{{#2}}}}{}#1}}}}
\newcommand{\la}{\langle}
\newcommand{\ra}{\rangle}
\newcommand{\ovl}[1]{\overline{#1}}
\newcommand{\Bak}[1]{\mathbf B_{#1}}
\newcommand{\bak}{\mathbf b}
\newcommand{\dg}[2]{\mathrm{ex}_{#1}#2}
\newcommand{\ord}[2]{\mathrm{ord}_{#1}#2}
\newcommand{\Ex}[1]{\!\!\uparrow^{#1}}
\newcommand{\Exx}[2]{(#1)\!\!\uparrow^{#2}}
\newcommand{\STDm}[1]{\mathrm{STD}({#1})}
\newcommand{\STD}[1]{{\rm STD}$({#1})$}

\newcommand{\s}[1]{{#1}^{\star}}

\newcommand{\beas}{\begin{eqnarray*}}
\newcommand{\eeas}{\end{eqnarray*}}
\newcommand{\bea}{\begin{eqnarray}}
\newcommand{\eea}{\end{eqnarray}}
\newcommand{\lcm}{\mathrm{lcm}}
\newcommand{\md}{\mathrm{mod\ }}
\newcommand{\expon}[2]{\mathrm{ex}_{#1}\left({#2}\right)}
\newcommand{\z}{\mathbf 0}
\newcommand{\rem}{\mathrm{rem}}

\newcommand{\rt}[1]{\mathrm{rem}\left(#1,\tau\right)}
\newcommand{\mulmi}[1]{\left\lfloor #1\right\rfloor_{\tau}}
\newcommand{\mulma}[1]{\left\lceil #1\right\rceil_{\tau}}
\newcommand{\mm}[1]{\,(\mathrm{mod}\ #1)}
\newcommand{\mv}[1]{\, \mathrm{mod}(#1)}

\newcommand{\pr}{{\sl Proof.\ }}
\newcommand{\bx}{\ \ \ $\Box$}
\newcommand{\ind}{\mathrm{ind}\,}
\newcommand{\Acal}{\mathcal{A}}
\newcommand{\Sal}{\mathcal{S}}
\newcommand{\Pal}{\mathcal{P}}
\newcommand{\Hal}{\mathcal{H}}
\newcommand{\Mal}{\mathcal{M}}
\newcommand{\Ral}{\mathcal{R}}
\newcommand{\Cal}{\mathcal{C}}
\newcommand{\ccirci}[2]{\mathcal{C}(#1)\big|_{#2}}
\newcommand{\rcirci}[2]{\mathcal{R}(#1)\big|^{#2}}
\newcommand{\Bfrak}{\mathfrak B}
\newcommand{\Bbol}{\mathbf B}

\newcommand{\mb}[1]{\mathbb{#1}}
\newcommand{\m}[1]{\mathfrak{#1}}

\newcommand{\ebxt}[2]{{#1}^{\boxtimes#2}}
\newcommand{\bxt}[2]{{#1}\boxtimes{#2}}
\newcommand{\BT}{\boxtimes}
\newcommand{\setp}{\{0,\dots,p-1\}}
\newcommand{\aut}[1]{\mathbf {Aut}(#1)}

\newcommand{\ED}{\end{document}}

\sloppy

\title[Isomorphisms of Additive CA on groups]{
Isomorphisms of Additive Cellular Automata on Finite Groups}

\author[V. Bulitko]{Valeriy Bulitko\\
\ \\
{\tiny Athabasca University}\\
{\tiny valeriyb@athabascau.ca}}

\vspace*{-2.5cm} \sloppy \thispagestyle{empty}

\DeclareGraphicsExtensions{.png,.gif,.jpg,.pdf}

\maketitle

\begin{abstract}
We study sources of isomorphisms of additive cellular automata on finite groups (called index-group). It is shown that many isomorphisms (called regular) of automata are reducible to the isomorphisms of
underlying algebraic structures (such as the index-group, monoid of automata rules, and its subgroup of reversible elements). However for some groups there exist not regular automata isomorphisms. A complete description of linear automorphisms  of the monoid
is obtained. These automorphisms cover the most part of all automata isomorphisms for small groups and are represented by reversible matrices $\Mal$ such that for any index-group circulant $\Cal$ the matrix $\Mal^{-1}\Cal\Mal$ is an index-group circulant.
\end{abstract}
\ \\
\indent\indent\indent{\bf Subj-class}: nlin. CG\\
\indent\indent\indent{\bf MSC-class}: 37B15, 68Q80

\footnotesize
\tableofcontents
\normalsize

\section{Introduction}

Classical cellular automata (CA) after S. Ulam and J. von Neumann \cite{Ulam-Neumann, Wolf} are defined on regular grids which are actually finite direct products of finite or infinite cyclic groups.
If to say about finite products of finite cyclic groups (tori ) then it appears that many different CA have actually the same behavior after renaming states. In other words, the state transition diagrams of many different automata are isomorphic.\footnote{For instance, there are 256 different rules for additive cellular automata (see definition further or \cite{linCA}) with two-state cells on 1-dimensional torus of size 8. Among them only 16 are essentially different.}

The grids have a certain system of symmetries that could be described by isomorphisms of the groups and these symmetries induce isomorphisms of behaviors of automata with different rules. Saying "behavior of an automaton" we mean the state transition diagram of the automaton.

However besides of the symmetries of grids there are other symmetries which influence automata behavior; for instance - symmetries of automata rules.

We study the question whether it is possible to derive all isomorphisms among state transition diagrams of CA from symmetries of underlying structures such that grids, sets of states of cells, rules, etc.

For that we first should determine these things in such a way that their symmetries were clearly seen. On the other hand our purpose is also to diversify the set of possible symmetries of the CA supports.
This is why we restrict ourselves with additive (i.e. linear homogeneous) automata whose sets of cell's states are finite fields because the set of rules of such automata has a clear algebraic structure. (The class of general ACA on grids is well known, see for instance \cite{linCA}.)  On the other hand we  diversify the set of supports via replacement of the classical grids with arbitrary (finite though) groups called further {\sl index groups}.

We consider additive cellular automata on finite groups as an appropriate frame to study the question because placing cells of an automaton in group elements and making rule applications such that hold the group symmetries we can observe more rich picture of the connection between the structures of the groups and isomorphisms of the automata than it can be seen for the particular case such as finite tori. About finite groups see for instance  \cite{Coxeter}.

One general expectation of course is that the most of isomorphisms (we call them {\sl regular}) can be reduced to the system of symmetries (isomorphisms) of an underlying algebraic structure, basic symmetries that should be determined. According to the description of the basic symmetries accepted in this work there are many groups such that all isomorphisms of the CA on them are reducible to the symmetries of the underlying algebraic structure (index group, monoid $\m M$, its subgroup of reversible elements, and other, see below). However for some groups there are isomorphisms of CA on them which cannot be reduced to the basic symmetries that we accept.

Section 2 deals with the definition of homogeneous linear automata on groups and the generalization of the notion of circulant \cite{Ryzhik}  onto groups and the group convolution \cite{Weis}: $\Cal(T)$ and $\BT$ playing central roles in the paper.

In section 3 we mainly study the automorphisms of a monoid $\m M$ created by automata rules and the operation $\BT$. First we study the contribution of the symmetries (index permutations) of the index group into the basic symmetries. The complete description of the class of index permutations for any given index group $\m g$  is a constructive relatively $\m g$. Then more wide class of linear automorphisms of $\m M$ is completely characterized. Both these classes are much easier to list than the complete set $\aut{\m M}$ of all automorphisms of $\m M$. Further we extend the class of the isomorphisms produced by $\aut{\m M}$ with some isomorphisms related to the group $\m G$ of all reversible elements of $\m M$ obtaining the set of regular isomorphisms of automata.

In section 4 discussing our construction we provide a proof that all isomorphisms among CA upon field $\m F_2$ on some cyclic index groups $\m c_q$ for whose orders $q$ the number 2 is a primitive root \cite {Rosen} modulo $q$ are regular. We conjecture this is true for all prime $q$ that the number 2 is a primitive root modulo $q$. Also several examples of index groups are given for which the class of all CA isomorphisms is wider than classes of regular isomorphisms.

\subsubsection{General denotations}
\ \\
\begin{itemize}
\item[a:] $\m g$ a finite group  $\{g_0,g_1,\dots,g_{n-1}\}$ with the unit $g_0=\mathfrak{1}$. We use denotation $\ovl g$ for inverse element to $g\in\m g$. Also we assume that $\m g$ is not a trivial, i.e. $n>1$. Note that the above enumeration of elements of $\m g$ instals a linear order on the group where $\m1$ is the first element. We call this group {\sl index-group}.

\item[b:] $p$ a prime number. We denote both kinds of multiplication: in the number field $\m F_p=\la\setp,0,1,+,\cdot\ra$ and in the group $\m g$ in the same way, - as usual, by simple concatenation of elements like  $kr,k,r\in\setp, gq,g,q\in\m g$.

\item[c:] $\m V=\{v|v:\m g\to\{0,\dots,p-1\}\}$ set of evaluations of elements of $\m g$ and $v(g),g\in\m g,$ is $g$-th component of $v$. Sometimes we use upper or low indices to select vector's components when vectors participate in matrix algebra operations as row-vector and column-vector respectively.

\item[d:] $g$th-{\sl constituent} $K[g]$ for which by the definition $K[g](g)=1$ and $K[g](g')=0$ for any $g'\neq g$.

\item[e:] For $v\in\m V$ let $v^{-}$ be an element of $\m V$ such that $v^{-}(g)=v(\ovl g),\ g\in\m g$.

\item[f:] $\m L=\la \m V,+,\cdot\ra$ a vector space on $\m V$ over field $\m F_p$ with vector's addition $(v+v')(g)=v(g)+v'(g)\mm p$ and multiplication of vectors by scalars $(k\cdot v)(g)=(kv)(g)=kv(g), k\in\setp,g\in\m g$.

\item[g:] The standard basis $\mathbf K=\{K[g_i]|g_i\in\m g\}$ for $\m L$.

\item[h:] If $P$ is $n\times n$ matrix, $P_i,P^j$, and $P_i^j$ denote relatively the $i$th row, $j$th column, and the element of $P$ at the intersection of $i$th row and $j$th column.

\end{itemize}

\section{ACA on groups}

For any vector $R\in\m V$ we define a {\sl additive linear cellular automaton on group} $\m g$ over field $F_p$ as a system whose {\sl states} consist $\m V$ and whose behavior $v^{[0]},v^{[1]},\dots,v^{[\tau]},\dots$ (where $\tau,\tau\in\mathbb Z^+,$ represents time\footnote{Usage of the parentheses $[,]$ in the denotation $v^{[\tau]}$ for a state at time $\tau$ is caused by a necessity to distinct $i$-th component $v^{i}$ of a row vector $v$ from the value of the vector at time $\tau$.}  and $v$ is an initial state) is defined by recursion:
\bea\label{def-aut}
v^{[0]}=v;\ v^{[\tau+1]}_f=\sum_gR(\ovl fg)v^{[\tau]}(g), f\in\m g.
\eea
This recursion reflects the fact that to calculate new state of cell $f$ we shift rule $R$ along the index-group by $f$. The shift could be expressed as $R(\ovl fg)$ or $R(g\ovl f)$ which are equivalent for commutative index-groups. We choose the first form.

Vector $R$ is called {\sl rule} of the automaton denoted $\Acal_{\m g}^{p}(R)$ of shorter as $\Acal(R)$ when index group $\m g$ and the field $\m F_p$ are fixed.
Note that in case when $\m g$ is a cartesian product of $k$ cyclic groups we deal with additive cellular automata on $k$-dimensional tori.

Let $R*v$ denotes the application of rule $R$ to state $v$ according to (\ref{def-aut}). Using this denotation we can rewrite (\ref{def-aut}) more concisely $v^{[\tau+1]}=R*v^{[\tau]}$.

{\sl State transition diagram} $\STDm R$ for automaton $\Acal (R)$ is defined as usual, i.e. this is a graph $\la\m V,\{(v,R*v)|v\in\m V\}\ra$. Automata with rules $R,T$ are {\sl isomorphic} if their diagrams $\STDm R, \STDm T$ are isomorphic; we denote the latter by $\STDm R\approx\STDm T$. Thus the set of all rules partitions on classes of isomorphisms whose quantity says how many essentially different ACA are there.

\subsection{Quantities of the classes of isomorphism for ACA on small groups}

Table~\ref{num} shows the numbers of classes of isomorphism for ACA with two-state cells on some small groups.\footnote{For some cyclic groups see also \cite{Bul2}.} Not only the number of classes depends of the structure of index group but some diagrams representing automata for a group do not appear among diagrams for another group of the same order.

\begin{table}[here]\label{num}
\caption{Number $|\m I|$ of classes of isomorphism for index groups of order $\le10$.
$\m c_m$ - cyclic group of order $m$;
$\m D_m$ - dihedral group of order $2m$; $\m Q$ - quaternion group (the denotations are from \cite{Coxeter}).}

\begin{tabular}{||l||r|r|r|r||}
\hline\hline
Index group $\m g$&Order&Commutative?&Number of Rules&\ \ $| \m I| $\ \\
\hline\hline
$\m c_2$&2&yes&4&4\\
\hline
$\m c_3$&3&yes&8&6\\
\hline
$\m c_4$&4&yes&16&8\\
\hline
$\m c_2\times\m c_2$&4&yes&16&6\\
\hline
$\m c_5$&5&yes&32&10\\
\hline
$\m c_6$&6&yes&64&24\\
\hline
$\m D_3$& 6&no&64&22\\
\hline
$\m c_7$&7&yes&128&12\\
\hline
$\m c_8$&8&yes&256&16\\
\hline
$\m c_2\times\m c_4$&8&yes&256&12\\
\hline
$\m c_2\times\m c_2\times\m c_2$&8&yes&256&8\\
\hline
$\m D_4$&8&no&256&14\\
\hline
$\m Q$&8&no&256&12\\
\hline
$\m c_9$&9&yes&512&42\\
\hline
$\m c_3\times\m c_3$&9&yes&512&30\\
\hline
$\m c_{10}$&10&yes&1024&40\\
\hline
$\m D_5$&10&no&1024&44\\
\hline\hline
\end{tabular}
\end{table}

\section{Algebraic structure of variety of HLCA on a group $\m g$ and its automorphisms}

\subsection{Group circulants}
\ \\

Using matrix-vector multiplications we normally admit that automata' states are columns whereas rules are rows.
That is why in these cases we notate components of a state as $v_g$ and rule's components as $R^g$.
The agreement allows rewrite the recursion above in a different way using the generalization of the standard \cite{Ryzhik} concept of circulant matrices.

Here {\sl circulant} is a matrix $\Cal$ whose $(g,f)$-components satisfy the condition $\Cal\big|_g^f=\Cal_{\m1}^{\ovl gf},f,g\in\m g$. Therefore if $\Cal_1=R$ we have $\Cal_g^f=R^{\ovl gf}$. Since in addition any circulant is defined by the first row ({\sl leader}) we use the denotation $\Cal(R)$ for a circulant $\Cal$ s.t. $\Cal_{\m 1}=R$.

\begin{lem}\label{crit-circ}
(i) $\Cal(H)$ is a circulant iff $\Cal(H)_{sg}^{sf}=\Cal(H)_{g}^{f}$.\\
(ii) $\Cal(T)\Cal(H)=\Cal(Q)$ where $Q^f=\sum_qT^{q}H^{\ovl qf}$.
\end{lem}

\pr (i) By the definition $\Cal(H)_{sg}^{sf}=H^{\ovl{sg}\,sf}=H^{\ovl{g}f}$. Now assume that a matrix $Q$ obeys the condition $Q_{sg}^{sf}=Q_{g}^{f}$ for all $f,g,s\in\m g$. Then setting $s=\ovl g$ we arrive at $Q_{g}^{f}= Q_{\m1}^{\ovl gf}$ which is a definition to $\Cal(Q_{\m 1})$.

(ii)  Indeed,
\bea
\left[\Cal(T)\Cal(H)\right]_{sg}^{sf}= \sum_u\Cal(T)_{sg}^u\Cal(H)_u^{sf}= \sum_uT^{\ovl{sg}\,u}H^{\ovl u\,sf}=_{\text{ setting $q:=\ovl{sg}\,u$}}=\notag\\
\sum_qT^qH^{\ovl{q}\,\ovl{sg}\,sf}=\sum_qT^qH^{\ovl{q}\,\ovl g\,f}=_{\text{ setting $\ovl u:=\ovl{q}\,\ovl{g}$}}=\label{g=1}\\
\sum_uT^{\ovl gu}H^{\ovl uf}=\sum_u\Cal(T)_g^u\Cal (H)_u^f =\left[\Cal(T)\Cal(H)\right]_g^f.\notag
\eea
Therefore by lemma~\ref{crit-circ} the product $\Cal(T)\Cal(H)$ is a row circulant and for its first row $Q$ it follows from~(\ref{g=1}) that $Q^f=\sum_qT^{q}H^{\ovl qf}$ when $g=\m1$.
\bx

Thus the dynamic equation for automata could be written using group circulant as $v^{t+1}=\Cal(R)v^t$.

Group circulants for cyclic groups are the standard circulant matrices \cite{Ryzhik}.

\subsection{Operations $\BT$ and $\bak_p$}

With the group convolution (see \cite{Weis}) $\boxtimes :\m V\times\m V\to\m V$ defined by
 \bea\label{BT-def}
 (v\boxtimes v')(g)=\underset{f}\sum v(f)v'(\ovl fg)\ v,v'\in\m V,\ g,g'\in\m g,
 \eea
the statement (ii) of lemma~\ref{crit-circ} could be expressed as
\begin{cor}
$\Cal(T)\Cal(H)=\Cal(T\BT H)$.
\end{cor}
The next consequence plays an important role in the following:
\begin{theo}\label{lem2}
$v^{[\tau+1]}=\bxt{v^{[\tau]}}{R^{-}}$.
\end{theo}
\pr $R*v=\bxt{v}{R^{-}}$ because from (\ref{def-aut}) we have $(R*v)(f)=\sum_gR(\ovl fg)v(g)=\sum_gv(g)R^-(\ovl gf)=(v\BT R^-)(f)$. \bx

We use also the derivative operation $\bak_p(T)=\underbrace{T\BT\dots\BT T}_{p\text{ terms}}$ which in case of finite abelian groups was studied in \cite{Bul1}.

Here are some properties of $\BT$ that we use.
Let $\m Z$ be the commutant of $\m g$. We call $A\in\m V$ as $\m Z$-{\sl correct} if  $\forall g,f[g\ovl f\in\m Z\implies A(g)=A(f)]$.

\begin{lem}\label{oper-bxt}
\begin{itemize}
\item[(i)] $\bxt{}{}$ is an associative operation.\\
\item[(ii)] $\bxt{K[\m1]}A=\bxt A{K[\m1]}=A$.\\
\item[(iii)] $\bxt{}{}$ is a linear operation on $\m L$ w.r.t. both operands:
\beas
A\BT(B+C)=A\BT B+A\BT C,\ \ (B+C)\BT A=B\BT A+C\BT A.
\eeas
\item[(iv)] $A^{-}\BT B^{-}=(B\BT A)^{-}$.\\
\item[(v)]   $A\BT B=B\BT A$ for any $A,B\in\m V$ such that at least one of them is $\m Z$-correct.
\item[(vi)] $K[f]\BT K[g]=K[fg]$ for any $f,g\in\m g,n\in\mathbb Z$.
\item[(vii)]
\beas
(K[g]\BT A)(f)= A(\ovl gf),\ \ \ \ (A\BT K[g])(f)= A(f\ovl g).
\eeas
In other words $K[g]\BT A=\Cal(A)_g$.
\end{itemize}
\end{lem}
\pr (i) For $A,B,C\in\m V,g\in\m g$ we have
\beas
[\bxt{(\bxt AB)}C](g)=\sum_f\left(\sum_rA(r)B(\ovl rf)\right)C(\ovl fg)= \sum_rA(r)\sum_fB(\ovl rf)C(\ovl fg)=_{h:=\ovl rf} \\ =\sum_rA(r)\left(\sum_{h}B(h)C(\ovl h\ovl rg)\right)= \sum_rA(r)(\bxt BC)(\ovl rg)=[\bxt A(\bxt BC](g).
\eeas
It is possible to replace bounded variable $f$ with $h$ since for each fixed value of $r$ mapping $f\mapsto \ovl rf$ is 1-1-mapping $\m g$ on $\m g$.\footnote{We use similar arguments in many places below.}\\
(ii) $(\bxt{K[\m1]}A)(g)=\sum_fK[\m 1](f)A(\ovl fg)=A(g)= \sum_fA(f)K[\m1](\ovl fg). $\\
(iii) Now
\beas
(A\BT(B+C))(g)=\sum_fA(f)(B+C)(g\ovl f)=\sum_fA(f)B(\ovl fg)+ \sum_fA(f)C(\ovl fg)=\\(A\BT B)(g)+(A\BT C)(g).
\eeas
The second identity has a similar proof.\\
(iv) $(R^{-}\BT Q^{-})(g)=\sum_{f}R^{-}(f)Q^{-}(\ovl fg)=\sum_{f}R(\ovl f)Q(\ovl gf)$. If we define $w=\ovl gf$ (for any fixed $g$ variable $w$ runs over $\m g$ while $f$ runs over $\m g$) then $\ovl f=\ovl w\,\ovl g$ and $\sum_{f}R(\ovl f)Q(\ovl gf)=\sum_{w}Q(w)R(\ovl w\,\ovl g)=(Q\BT R)(\ovl g)=(Q\BT R)^{-}(g)$. \bx\\
(v) By definition we can write:
\beas
(\bxt AB)(g)=\sum_fA(f)B(\ovl fg)=\sum_rB(r)A(g\ovl r)\\
(\bxt BA)(g)=\sum_rB(r)A(\ovl rg)=\sum_fA(f)B(g\ovl f).
\eeas
Now if for instance $B$ is $\m Z$-correct then $B(\ovl fg)=B(g\ovl f)$ for all $g,f\in\m g$. This is because $g\ovl f\,\ovl{\ovl fg}$ is a commutator. Hence $(\bxt AB)(g)=(\bxt BA)(g)$.\\
(vi)
\beas
(K[f]\BT K[g])(r)=\sum_sK[f](s) K[g](\ovl sr) =K[g](\ovl fr)=
\begin{cases}
1,& \ovl fr=g\\
0& \text{otherwise}.
\end{cases}
\eeas
Since $\ovl fr=g$ means $r=fg$ we get what we need.\\
(vii)
\beas
(K[g]\BT A)(f)=\sum_sK[g](s)A(\ovl sf)=_{\text{since }K[g]=0\text{ if }g\neq s}=A(\ovl gf).\\
(A\BT K[g])(f)=\sum_sA(s)K[g](\ovl sf)=_{\text{since }K[g]=0\text{ if }g\neq \ovl sf}=A(f\ovl g).
\eeas
\bx

\subsection{Monoid $\m M$}

As $\BT$ is an associative binary operation on $\m V$ the structure $\m M=\la\m V,\boxtimes,K[\mathfrak1]\ra$ is a monoid \cite{Clifford} with the unit $K[\mathfrak1]$.

We denote $\m G$ the subgroup of the monoid consisting of all its reversible elements (i.e. such $A\in\m V$ that there exists $B\in\m V$ obeying $A\BT B=B\BT A=K[\m1]$).
\begin{lem}\label{revers-circ}
(i) $A\in\m M$ is reversible iff $\Cal(A)$ is a non singular matrix.\\
(ii) Mapping $g\mapsto K[g]$ is isomorphism between $\m g$ and subgroup $\m K$ of $\m G$ generated by vectors $\{K[g]|g\in\m g\}$.
\end{lem}

\pr  (i) Indeed, if for a vector $B$ we have $A\BT B=K[\m1]$ then $\Cal(A)\Cal(B)$ is equal to the identity matrix $\Cal(K[\m1])$ and therefore $\Cal(A)$ is non singular.

Vice versa, if $\Cal(A)$ is non singular then a unique matrix $\Mal$ exists obeying $\Mal\Cal(A)=\Cal(K[\m1])$. Hence $\Mal$ also is non singular. Let us show that $M$ is a circulant. For that let us prove that an inverse matrix of a circulant is a circulant, i.e. $[\Cal(A)]^{-1}|_g^f=[\Cal(A)]^{-1}|_{\m1}^{\ovl gf}$.

Consider $\Mal\Cal(A)=\Cal(K[\m1])$ as an equation's system for $\Mal$ which in terms of elements looks like
\beas
\sum_s\Mal_g^sA^{\ovl sf}=K[\m1]^{\ovl gf}.
\eeas
Let search $\Mal$ in form of a circulant $\Cal(M)$. Then the system above can be rewritten as
\beas
\sum_sM^{\ovl gs}A^{\ovl sf}=K[\m1]^{\ovl gf}.
\eeas
Since the matrix of this system of linear equations for unknown numbers $M^i,i\in\m g,$ is $\Cal(A)$ and $|\Cal(A)|\neq0$, a unique solution $M$ exists to this system. Hence $\Mal=\Cal(M)$ and $\Cal(M)\Cal(A)=\Cal(K[\m1])$. Finally $M\BT A=K[\m1]$. The similar reasonings yield $A\BT M=K[\m1]$. Therefore vector $A$ is reversible in $\m M$.

(ii) follows from lemma~\ref{oper-bxt} (i),(ii),(vi).
\bx

The monoid $\m M$ and linear space $\m L$ are subsystems of an associative algebra $\mathbf  A=\la\m V,+,\BT \ra$ over field $\m F_p$. Clearly $\m A$ is the enveloping algebra of Lie algebra $L(\mathbf A)$ after introducing Lie bracket as follows:
\beas
[A,B]=A\BT B- B\BT A.
\eeas
\begin{lem}
If $A$ or $B$ is $\m Z$-correct then $[A,B]=[B,A]=0$.
\end{lem}

\pr Indeed,
\beas
[A,B](f)=(A\BT B-B\BT A)(f)= \sum_gA(g)B(\ovl gf)-\sum_rB(r)A(\ovl rf)=_{\text{setting }g=\ovl rf}=\\ \sum_gA(g)B(\ovl gf)-\sum_gA(g)B(f\ovl g).
\eeas
Hence if $B$ is $\m Z$-correct, then $[A,B]=\mathbf0$. The same is true for $\m Z$-correct $A$ instead of $B$ because $[A,B]=-[B,A]$.
\bx

The monoid is the basic algebraic structure on the set $\m V$ of all states and rules of HLCA on $\m g$ relatively to the question that we study in this paper. Indeed, state transition diagram for a rule $w^-\in\m V$ is completely defined in terms of $\m M$ as a graph $\la\m V,\{(v,v\BT w)|v\in\m V\}\ra$. Therefore the more basic is an algebraic structure on HLCA whose automorphisms generate isomorphisms of diagrams the more complete set of isomorphisms among the diagrams can be revealed. On the other hand completions $\m M$ with additional operations lead to algebraic structures whose automorphisms are more specific but admit often more simple description. The next section illustrates this.

We call group ${\m g}^{\circ} (\m G^{\circ})$ with the group multiplication $\circ(x,y)$ co-group for $\m g$ if it consists of the same elements as $\m g\ (\m G)$ and for all group elements $f,q$ it holds $[\circ(f,q)=f\circ q=qf]$ where $qf$ if the group multiplication in $\m g(\m G)$. Thus the table of multiplication of $\m g^{\circ}\ (\m G^{\circ})$ is a transposed table of multiplication of $\m g\ (\m G)$.

{\sl Reflection on $\m g\ (\m G)$} is a mapping $\rho:\m g\to\m g\  (\varrho:\m G\to\m G)$ such that  $\rho(f)=\ovl f, f\in\m g\  (\varrho(f)=\ovl f, f\in\m G)$.

Similarly we call {\sl co-monoid} to $\m M$ a monoid $\m M^{\circ}=\la\m V,\BT^{\circ}, K[\m1]\ra$ where $v\BT^{\circ}w=w\BT v$, and {\sl reflection on $\m M$} is a mapping $\rho:\m V\to\m V$ such that  $\rho(v)=v^-, v\in\m V$.

\begin{lem}\label{refl}
1). The reflection on $\m g$ ($\m G,\m M$) is a natural isomorphism between given index-group $\m g$ (group $\m G$, monoid $\m M$) and the co-group $\m g^{\circ}$ (co-group $\m G$, co-monoid $\m M^{\circ}$).\\
2). Any isomorphism $\psi:\m g\to\m g^{\circ}$ ($\psi:\m M\to\m M^{\circ}$) is a composition of an automorphism $\varphi$ of  $\m g$  ($\m M$) and the reflection.
\end{lem}

\pr  Clearly the reflections are 1-1-mappings.

1) For index-group $\m g$: $\rho(x)=\ovl x, \rho(\m1)=\m1, \rho(xy)=\ovl{xy}=\ovl y\ovl x=\ovl x\circ\ovl y=\rho(x)\circ\rho(y)$.

For the monoid $\m M$: $\rho(R)=R^-, \rho(K[\m1])=K[\m1]^-=K[\m1]$. Then,
according to lemma~\ref{oper-bxt} (iv) we have
$\rho(R\BT T)=(R\BT T)^-=T^-\BT R^-=\rho (T)\BT\rho (R)=\rho(R)\BT^{\circ}\rho(T)$.

2) If $\psi$ is an isomorphism then $\rho\psi$ is an automorphism $\varphi$. Therefore $\psi=\rho\varphi$ because the reflection is idempotent.
\bx

\begin{theo}\label{aut->iso}
If $\varphi:\m M\to\m M$ is an automorphism of the monoid and $T,R$ are automata rules such that $T=\varphi(R)$ then $\STDm T\approx\STDm R$ and $\STDm{T^-}\approx\STDm{R^-}$.
\end{theo}

\pr
From theorem~\ref{lem2} we have $R*v=v\BT R^-$. On the other hand lemma~\ref{oper-bxt} (iv) yields $(v\BT R^-)^-=R^{--}\BT v^-$. Since the operation $(\cdot)^-$ is idempotent we obtain $(v\BT R^-)^-=R\BT v^-$. Thus
\bea\label{forms}
w=R*v \iff w=v\BT R^- \iff w^-=R\BT v^-
\eea
The similar chain of the equivalents holds for $T$ as well.
Now from $\varphi(w^-)=\varphi(R)\BT\varphi(v^-)=T\BT\varphi(v^-)$ and the fact that the mapping $(\cdot)^-$ is 1-1-mapping on $\m V$ the isomorphism of $\STDm R,\STDm T$ follows.

On the other hand since $R^{--}=R$ we see that $(v,w)$ is an edge of $\STDm R^-$ iff $w=v\BT R$. As $T=\varphi(R)$ given we arrive at $\varphi(w)=\varphi(v)\BT T$. This means that $\STDm{T^-}\approx\STDm{R^-}$.
\bx\footnote{In other words we apply here $\varphi(R)=T$ to $R\BT v^-$ and $v\BT R$. The first leads to isomorphism of the diagrams of $R,T$ and the second - to $\STDm{T^-}\approx\STDm{R^-}$.}

From here it does not follow yet that $\STDm R\approx\STDm{R^-}$ whereas examples (with non-abelian index-groups) show that $\STDm R\approx\STDm {R^-}$. In abelian case we have

\begin{theo}
If $\m g$ is commutative group then $\varphi: R\mapsto R^-,R\in \m V,$ is an automorphism of $\m M$. Therefore $\STDm R\approx\STDm{R^-}$ for any $R\in\m V$.
\end{theo}
\footnote{This is also a consequence of a theorem about functional index-permutations (see below).}
\pr The commutant $\m Z$ of any commutative group is consists of one element that is the unit of the group. Therefore any $v\in\m V$ is $\m Z$-correct and from lemma~\ref{oper-bxt}(v)  it follows that $\BT$ is commutative for commutative $\m g$. Therefore (iv) from lemma~\ref{oper-bxt} looks like $(v\BT u)^-=v^-\BT u^-$. This proves that $\varphi: R\mapsto R^-,R\in \m V,$ is an automorphism of $\m M$ translating any rule $R$ into $R^-$.
Hence $\STDm R\approx\STDm{R^-}$. \bx

\subsection{Index-permutations}

One class of automorphisms of the monoid, a class that has a simple description is defined by automorphisms of the index-group.

We call {\sl index-permutation} any permutation $\theta:\m g\to \m g$ of the index-group $\m g$. Any index-permutation $\theta$ generates 1-1-mapping $\Theta:\m V\to\m V$ by a rule $\Theta(v)|_f:=v|_{\theta(f)}$. We say that index-permutation $\theta$ is {\sl functional index-permutation} if for any rule $R\in\m V$ there exists a rule $T$ such that $T*v=\Theta^{-1}(R*\Theta(v))$ for any state $v\in\m V$. In this case $R,T$ are obviously isomorphic.

\begin{theo}
1. An index-permutation $\theta$ is functional iff
$\ovl{\theta(\m1)}\theta(\cdot)$ is an automorphism of $\m g$.\\
2. If $\theta$ is a functional index-permutation and rule $T$ satisfies $T*v=\Theta^{-1}(R*\Theta(v)),v\in\m V,$ then $T_{\ovl{\theta(\m1)}\theta(h)}=R_h,h\in\m g$.
\end{theo}

\pr
Assume $\Theta(T*v)=R*\Theta(v), v\in\m V$. Then for each $f\in\m g$
\beas
\sum_hR_{\ovl {f}\,h}v_{\theta(h)}=\sum_hT_{\ovl{\theta(f)}\,h}v_h=
\sum_hT_{\ovl{\theta(f)}\,\theta(h)}v_{\theta(h)}.
\eeas
Since $v$ runs over $\m V$ it must hold $R_{\ovl {f}\,h}=T_{\ovl{\theta(f)}\,\theta(h)}, h,f\in\m g$. From here we deduce the condition that the product $\ovl{\theta(f)}\,\theta(h)$ depends only on $\ovl {f}h$. Let a function $\varphi:\{g_0,\dots,g_{n-1}\}\to\{g_0,\dots,g_{n-1}\}$  satisfies
\bea\label{theta}
\ovl{\theta(f)}\,\theta(h)=\varphi(\ovl {f}h),\ h,f\in\m g,
\eea
and denote $a=\theta(\m1)$.

Setting $f:=\m1$ we get $\theta(h)=a\varphi(h)$ and conclude that $\varphi$ is 1-1-mapping on $\m g$. Yet if to substitute $h:=\m1$ we arrive at $\ovl{\theta(f)}a=\varphi(\ovl f)$ and  since $f,h$ any elements of $\m g$ we conclude $\varphi(\ovl x)=\ovl{\varphi(x)}, x\in\m g$. On the other hand from (\ref{theta}) setting $h=f=\m1$ we get $\varphi(\m1)=\m1$ and substituting there $a\varphi(h),a\varphi(f)$ instead of $\theta(h),\theta(f)$ relatively we arrive at $\varphi(\ovl {f}\,h)=\varphi(\ovl f)\phi(h)$.
Thus we obtain the following equations for $\varphi$:
\bea
\begin{cases}\label{ip8}
\forall x,y\in\{g_0,\dots,g_{n-1}\}[x\neq y\implies\varphi(x)\neq\varphi(y)],\\
\varphi(\m1)=\m1,\\
\varphi(\ovl x)=\ovl{\varphi(x)},\\
\varphi(xy)=\varphi(x)\phi(y).
\end{cases}
\eea
If $\varphi$ is being considering as a mapping from $\m g$ into $\m g$, then $\varphi$ is an automorphism.

Simultaneously we proved the second statement because from $R_{\ovl {f}\,h}=T_{\ovl{\theta(f)}\,\theta(h)}$ the equation $T_{\ovl{\theta(\m1)}\theta(h)}=R_h$ follows if to set $f:=\m1$.

Vice versa, an automorphism $\phi$ of $\m g$ and $a\in\m g$ given, let us define an index permutation $\theta(\cdot)$ as $a\phi(\cdot)$ and let $R$ be any rule. First if $u=\ovl{\theta^{-1}(f)}$ then $f=\theta(\ovl u)=a\phi(\ovl u)$ and $u=\ovl{\phi^{-1}(\ovl af)}$. Hence
\beas
\Theta^{-1}(R*\Theta(v))|_f=(R*\Theta(v))|_{\theta^{-1}(f)}= \sum_hR_{\ovl{\theta^{-1}(f)}\,h}v_{a\phi(h)}= \sum_hR_{\ovl{\phi^{-1}(\ovl a\,f)}\,h}v_{a\phi(h)}=_{q:=a\phi(h)}\\
\sum_qR_{\ovl{\phi^{-1}(\ovl a\,f)}\,\phi^{-1}(\ovl a\,q)}v_{q}.
\eeas
And because $\phi^{-1}$ also is an automorphism of $\m g$ we can continue as
\beas
\sum_qR_{\ovl{\phi^{-1}(\ovl a\,f)}\,\phi^{-1}(\ovl a\,q)}v_{q}=
\sum_qR_{(\ovl{\phi^{-1}(f)})\,(\ovl{\phi^{-1}(\ovl a)})\,\phi^{-1}(\ovl a)\phi^{-1}(q)}v_{q} =\sum_qR_{\phi^{-1}(\ovl f)\,\phi^{-1}(q)}v_{q}.
\eeas
It remains to note that $\phi^{-1}(\ovl f)\,\phi^{-1}(q)=\phi^{-1}(\ovl f\,q)$ to conclude that it is possible to define a rule $T$ as $T_{\ovl f\,g}:=R_{\phi^{-1}(\ovl f\,q)}$.
\bx

{\sl Note:} from here the result $T$ of application of $\theta$ to $R$ is defined by $T_q=R_{\theta^{-1}(\theta(\m1)q)}$.\bx\footnote{Check this on examples.}

\begin{theo}\label{ind-aut-mon}
If $\theta$ is an automorphism of index-group $\m g$ then $\Theta:\m V\to\m V$ is an automorphism of monoid $\m M$.
\end{theo}

\pr
\beas
\Theta(R\BT T)|_f=\sum_hR_hT_{h\theta(f)}=_{\theta(q):=h}= \sum_{\theta(h)}R_{\theta(h)}T_{\ovl{\theta(h)}\theta(f)}= \sum_{h}R_{\theta(h)}T_{\theta(\ovl{h}\,f)}= \\
\sum_{h}\Theta(R)_{h}\Theta(T)_{\ovl{h}\,f}=(\Theta(R)\BT\Theta(T))|_f
\eeas
\bx

\begin{cor}
If index-group $g$ is commutative then $\STDm{R}\approx\STDm{R^-}$ for each rule $R\in \m V$.
\end{cor}

\pr
For abelian $\m g$ the mapping $\theta:g\mapsto\ovl g$ is an automorphism. Hence by the theorem~\ref{ind-aut-mon} $\Theta$ induced by $\theta$ is an automorphism of the monoid $\m M$. On the other hand, $\Theta(R)_f=R_{\theta(f)}=R_{\ovl{f}},f\in\m g$. Therefore $\Theta(R)=R^-$.
\bx

Examples show that at least for some non-commutative rules  $\STDm{R}\approx\STDm{R^-}$ for each rule $R\in \m V$. Is this true or not for all index group could not be solved on the basis of index-permutations only since $\theta: g\mapsto\ovl g, g\in\m g,$ is not an automorphism for any non-commutative group $\m g$.

\subsection{Linear automorphisms of $\m M$}

We call an automorphism $\varphi:\m M\to\varphi M$ {\sl linear} automorphism if $\varphi(T+L)=\varphi(T)+\varphi(L),T,L\in\m V$.

It is trivially follows from here that
\begin{theo}\label{ind-lin}
Any automorphism $\Theta$ of $\m M$ generated by an automorphism of the index-group $\m g$ of the monoid $\m M$ is a linear automorphism of $\m M$.
\end{theo}

\pr
By the definition $\Theta(T+L)|_f=(T+L)|_{\theta(f)}=T_{\theta(f)}+L_{\theta(f)} = \Theta(T)|_f+\Theta(L)|_f= (\Theta(T)+\Theta(L))|_f$.
\bx

Now let us find general characteristics of linear isomorphisms.

\begin{lem}\label{autom->H}
A linear automorphism $\varphi$ given, let $H[g]$ be $\varphi(K[g]), g \in \m g$.
Then:\\
(1) Mapping $\psi:g\mapsto H[g], g\in\m g,$ is an injection of $\m g$ into the group $\m G$ consisting of all reversible elements of the monoid $\m M$.\\
(2) Rank of the system $\{H[g]|g\in\m g\}$ is equal to the order $|\m g|$ of the index group.
\end{lem}

\pr
(1) First of all $H[\m1]=K[\m1]$ because $\varphi$ is an automorphism of $\m M$. Then by the same reason and definition of $H$ we have $H[g]\BT H[f]=\varphi(K[g])\BT\varphi(K[f])=\varphi(K[g]\BT K[f]) =_{\text{by lemma~\ref{oper-bxt}(iv)}}=\varphi(K[gf])=H[fg]$. That means by the way that all $H[g], g\in\m g,$ are reversible in the monoid. In addition $\varphi$ is 1-1-mapping. Therefore $\psi:g\mapsto H[g]$ is an injection of the group $\m g$ into the monoid. Therefore all $H[g]$ should be reversible elements of the monoid $\m M$. Since all reversible elements of $\m M$ create a subgroup ($\m G$) of the monoid we deal with an injection $\m g$ into $\m G$.

(2) The system $\{H[g]|g\in\m g\}$ of vectors is linearly independent because the matrix $\Phi$ defined as $\Phi_g=\varphi(K[g]),g\in\m g,$ should be reversible as the matrix representation of the automorphism $\varphi$ considered as a linear operator on $\m L$.
\bx

Any injection $\psi$ of $\m g$ into $\m G$ obeying the condition that the system $\{\psi(g)|g\in\m g\}$ is linearly independent in $\m L$ we call {\sl non-singular $\m g$-injection}.

\begin{cor}
For any linear automorphism $\varphi$ of the monoid the mapping $\psi:\m g\to\m G$ defined by $\psi:g\mapsto \varphi(g)$ is a non-singular $\m g$-injection.
\end{cor}

If $\Mal$ is a non-singular matrix of size $|\m g|\times|\m g|$ and $\Cal(L)\Mal=\Mal\,\Cal(T)$ then rules $L,T$ produce isomorphic state transition diagrams. We call (see [2]) a non-singular matrix {\sl universal} if it commutes with set $\{\Cal(T)|T\in\m V\}$, i.e. $\forall T\in\m V\exists L\in\m V[\Cal(T)\Mal=\Mal\,\Cal(L)]$.
Obviously, for any matrix $\Pal$ of size $|\m g|\times|\m g|$, the matrices $\Mal,\Mal^{-1}$ are or are not universal simultaneously.

\begin{theo}
For any non-singular $\m g$-injection $\psi$ the matrix $\Psi$ defined as $\Psi_g=\psi(g),g\in \m g,$ is universal and in addition $\Psi_{\m1}=K[\m1]$.
\end{theo}

\pr The equality $\Psi_{\m1}=K[\m1]$ follows directly from the fact that $\psi(\m1)$ should be the unit $K[\m1]$ of the monoid.
By the definition $\psi(g)=K[g]\Psi, g\in \m g$. Then
\beas
\Psi\Cal(\psi(g))=\Cal(K[g])\Psi.
\eeas
Indeed,
\beas
\left[\Cal(K[g])\Psi\right]_f^h=\sum_s\Cal(K[g])|_f^s\Psi_s^h= \sum_sK[g]^{\ovl fs}\Psi_s^h=\Psi_{fg}^h,
\eeas
where the latter equality is caused by the fact that $K[g]^{\ovl fs}=0$ if $fg\neq s$.

On the other hand
\beas
\left[\Psi\Cal(\psi(g))\right]_f^h=\sum_s\Psi_f^s\Cal(\psi(g))_s^h= \sum_s\Psi_f^s\psi(g)|^{\ovl sh}= \sum_s\Psi_f^s\left({K[g]\Psi} \right)^{\ovl sh}=\\
\sum_s\Psi_f^s\sum_tK[g]^t\Psi_t^{\ovl sh}=\sum_s\Psi_f^s\Psi_g^{\ovl sh}=[\Psi_f\BT\Phi_g]^h=\Psi_{fg}^h
\eeas

Let $T\in\m V$ be a given rule. Then obviously
\beas
\Cal(T)=\sum_{g\in\m g}a_g\Cal(K[g]), a_g\in\m F_p.
\eeas
From here and proved above:
\beas
\Cal(T)\Psi=\sum_ga_g\Cal(K[g])\Psi=\sum_ga_g\Phi\Cal(\psi(g))= \Psi\left(\sum_ga_g\Cal(\psi(g))\right)=\Psi\Cal(L)
\eeas
where $L=\sum_ga_g\psi(g)$.
This proves that for any $T\in\m V$ there exists $L\in\m V$ such that $\Cal(T)\Psi=\Psi\,\Cal(L)$.
Also because $\psi$ is a non-singular $\m g$-injection the rank of $\Psi$ is equal to $|\m g|$, i.e. $\Psi$ is non-singular matrix.
Thus $\Psi$ in a universal matrix.
\bx

\begin{theo}
If  $\Mal$ is a universal matrix obeying $\Mal_{\m1}=K[\m1]$ then the transformation $\varphi:T\mapsto T\Mal, T\in\m V,$ is a linear automorphism of $\m M$.
\end{theo}

\pr In first, $\varphi$ is 1-1-mapping because $\Mal$ is non-singular. The linearity of $\varphi$ is also obvious and it remains only to prove that $\varphi(T\BT L)=(\varphi(T))\BT(\varphi(L))$ for all $T,L\in\m V$.

The remaining part of a proof of this theorem consists of several lemmas.

We say that ${H[g]}\in\m V$ is a {\sl  $g$-response} of a matrix $\Mal$ if $\Cal(K[g])\Mal=\Mal\Cal(H[g])$.
Since $\Mal$ is non-singular then the responses $H[g]$ are defined uniquely.

Evidently, $H[\m1]=K[\m1]$ because $C(K[\m1])$ is the identity matrix $\mathcal I$.

\begin{lem}\label{crit-P}
$\Cal(K[g])\Mal=\Mal\Cal(H[g])\iff\forall h[\Mal_{hg}=\bxt{\Mal_h}{H[g]}],\ \  g\in\m g$.
\end{lem}

\pr
For the system of responses $\{H[g]|g\in\m g\}\subseteq\m V$ of $\Mal$ it holds
\bea\label{recH}
\forall g\in\m g[\Mal\Cal(H[g])=\Cal(K[g])\Mal \iff
\forall  h,q\in\m g(\Mal^q_{hg}=\sum_u\Mal^u_h{H[g]}^{\ovl uq})],
\eea
Indeed,
\beas
\Mal^q_{hg}=\sum_uK[g]^{\ovl hu}\Mal^q_u=\sum_u\Cal(K[g])_h^u\Mal^q_u=(\Cal(K[g])\Mal)_h^q=\\
(\Mal\Cal({H[g]}))_h^q=
\sum_u\Mal^u_h\Cal({H[g]})_u^q=\sum_u\Mal^u_h{H[g]}^{\ovl uq}.
\eeas
Then we can rewrite (\ref{recH}) in form
\beas
\Mal_{hg}(q)=\sum_u\Mal_h(u)H[g](\ovl uq),\ h,q\in\m g.
\eeas
Hence for rows $\Mal_h,\Mal_{hg}$ of the universal matrix $\Mal$ and response $H[g]$ we get the equation $\Mal_{hg}=\Mal_h\BT H[g]$. \bx

For any reversible element  $A$ of the monoid $\m M$ we denote as $\ovl A$ a vector $X\in\m V$ such that $A\BT X=X\BT A=K[1]$ reserving the denotation $(\cdot)^{-1}$ for reverse matrix.

\begin{lem}
All responses $H[g],g\in\m g,$ of any universal matrix $\Mal$ are reversible in the monoid. In addition
$\ovl{H[g]}=H[\ovl g]$ and $\Cal(H[\ovl g])=\Cal^{-1}(H[g])$ .
\end{lem}

\pr
According to lemma~\ref{revers-circ} (i)  for any $A\in\m V$ the matrix $\Cal(A)$ is reversible in $\m L$ iff $A$ is reversible in $\m M$. And because the uniqueness of the reverse matrix (if it exists), when $\ovl A$ exists then $\Cal(\ovl A)\Cal(A)=\Cal(K[\m1])$ and therefore $\Cal^{-1}(A)=\Cal(\ovl A)$.
Let $\Mal$ be a universal matrix and $\Mal\Cal(H[g])=\Cal(K[g])\Mal,g\in\m g$. Since $\Cal(K[g])$ is non singular matrix $\Cal(H[g])$ also is non singular. Therefore $\Cal^{-1}(H[g])\Mal^{-1}=\Mal^{-1}\Cal^{-1}(K[g])$ and from here $\Mal\Cal^{-1}(H[g])=\Cal^{-1}(K[g])\Mal$.

On the other hand $\Cal^{-1}(K[g])$ is a circulant $\Cal(A)$ (as it is proved before), and $\Cal(K[\m1])=\Cal(A)\Cal(K[g])=\Cal(A\BT K[g])$. Hence $A\BT K[g]=\m1$. From lemma~\ref{oper-bxt}(vii) we get $A(u)\neq0\iff u=\ovl g$ and $A({\ovl g})=\m1$, i.e. $A=K[\ovl g]$.
\bx

\begin{lem}\label{gen}
Let $\Hal_g=H[g],g\in\m g,$ be the matrix compiled from the responses for a universal matrix $\Mal$. We call $\Hal$ a {\sl response matrix} for $\Mal$.\\
(i) If $f,h_1,\dots,h_k\in\m g$, and $f=h_1\dots h_k$ then $\Mal_f=\Mal_{\m1}\BT H[h_1]\BT \dots\BT H[h_k]$.\\
(ii) Each row $\Mal_s,s\neq\m1,$ is the result of an application of operation $\BT$ to $\Mal_{\m1}$ and some elements of $\mathbf H=\{H[g]|g\in\mathbf G\}$ where $\mathbf G$ is any fixed system of generators for $\m g$.\\
(iii) $H[f]\BT H[g]=H[fg]$ for each $f,g\in\m g$.\\
(iv) The system or rows of the response matrix $\Hal$ is linearly independent.
\end{lem}

\pr (i) From corollary ~\ref{crit-P} we have $\Mal_{h_1}=\Mal_{\m1}\BT H[h_1]$. Then $\Mal_{h_1h_2}=\Mal_{h_1}\BT H[h_2]$ and so on. Finally $\Mal_{f}=\Mal_{h_1\dots h_k}=\Mal_{\m1}\BT H[h_1]\BT \dots\BT H[h_k]$.

(ii) Fix any system $\mathbf G$ of generators for $\m g$. Since each element $s$ of the group is a product $g_1...g_k$ of some its generators from $\mathbf G$ and their inverse (where factors could repeat), we get from (i) that $\Mal_{s}= \bxt{\Mal_{\m1}}{H[g_1]}\bxt{}{\dots}\bxt{}{H[g_k]}$.

(iii)
From lemma~\ref{oper-bxt}(vi) we have $\Cal(K[f])\Cal(K[g])=\Cal(K[fg])$. Now
\beas
\Mal\Cal(H[f])\Cal(H[g])=\Cal(K[f])\Mal\Cal(H[g])=\Cal(K[f]) \Cal(K[g])\Mal=\Cal(K[fg])\Mal=\Mal\Cal(H[fg]).
\eeas
Hence $\Mal\{\Cal(H[f])\Cal(H[g])-\Cal(H[fg])\}=\mathbf0$. But $\Mal$ is non singular matrix.

(iv) Since $\Mal$ is non-singular matrix its system $\{\Mal_g|g\in\m g\}$ of rows is linearly independent.
On the other hand $\Mal_g=\Mal_{\m1}\BT H[g]$. From lemma~\ref{oper-bxt}(iii)
\beas
\sum_ga_g\Mal_g=\Mal_{\m1}\BT\sum_ga_gH[g].
\eeas
Hence if $\sum_ga_gH[g]=\mathbf0$ for some collection of $a_g\in \m F_p,g\in\m g,$ then $\sum_ga_g\Mal_g=\mathbf 0$.  \bx

\begin{lem}
For a universal matrix $\Mal$ and its response matrix $\Hal$ it holds
\bea\label{univ-resp}
\Mal\Cal(T\Hal)=\Cal(T)\Mal, \ \ T\in\m V.
\eea
\end{lem}

\pr
We have $T=\sum_gT^gK[g]$ and from here $T\Hal=\sum_gT^gK[g]\Hal=\sum_gT^g\Hal_g=\sum_gT^gH[g]$.

On the other hand
\beas
\Cal(T)\Mal=\Cal\left(\sum_gT^gK[g]\right)\Mal= \sum_gT^g\Cal(K[g])\Mal= \\\sum_gT^g\Mal\Cal(H[g])=
\Mal\Cal\left(\sum_gT^gH[g]\right)=\Mal\Cal(T\Hal).
\eeas
\bx

To finish the proof of the theorem we note that
$\Cal((T\BT L)\Hal)=\Mal^{-1}\Cal(T\BT L)\Mal=\Mal^{-1}\Cal(T)\Mal\Mal^{-1}\Cal(L)\Mal=\\
\Cal(T\Hal)\Cal(L\Hal)=\Cal((T\Hal)\BT(L\Hal))$
or $(T\BT L)\Hal=(T\Hal)\BT(L\Hal)$.

Finally when $\Mal_{\m1}=K[1]$ it coincides with its response matrix, that is $\Mal=\Hal$, and we arrive at $(T\BT L)\Mal=(T\Mal)\BT(L\Mal)$. \bx

Let for matrix $\mathcal M$ the matrix $\Mal^-$ be defined by the condition
$(\Mal^-)_g=(\Mal_g)^-,g\in\m g$.

\begin{lem}\label{reversible}
If $\Hal$ is a response matrix for $\Mal$ then $\Mal^-=\Hal^-\,\,\Cal (\Mal_{\m1}^-)$.
\end{lem}

\pr
We should prove $\Mal^-_g|^f=\sum_u\Hal^-_g|^u\,\Cal (\Mal_{\m1}^-)|_u^f$:
\beas
\sum_u\Hal^-_g|^u\,\Cal (\Mal_1^-)|_u^f=\sum_uH[g]^{\ovl u}\Mal_1^{\ovl{\ovl uf}}=_{s:=\ovl f\,u}\ \sum_s\Mal_1^{s}H[g]^{\ovl s\ovl f}=(\Mal_1\BT H[g])^{\ovl f}=\Mal_g^-|^f
\eeas
\bx

\begin{cor}
(i) Any two universal matrix $\Mal,\Mal'$ having the same response matrix $\Hal$ are equivalent in the meaning that
$\Mal\Cal(L)=\Cal(T)\Mal\iff \Mal'\Cal(L)=\Cal(T)\Mal', T,L\in\m V.$\\
(ii) $\Mal_{\m1}$ is a reversible in the monoid for any universal matrix $\Mal$.\\
(iii) If $A$ is any reversible element of \ $\m M$ and $\Mal$ is a  universal matrix with a response matrix $\Hal$, then matrix  $\Mal'$ defined as $\Mal'_g=A\BT\Mal_g$ is also a universal matrix with the same response matrix $\Hal$.
\end{cor}

\pr
(i) directly follows from (\ref{univ-resp}).

(ii) As any universal matrix is reversible it follows from lemma~\ref{reversible} that the matrix $\Cal(\Mal^-_{\m1})$ is reversible.  By lemma~\ref{revers-circ} (i) then $\Mal_{\m1}^-$ is a reversible element in the monoid, i.e. there exists an element $X\in\m V$ such that $X\BT\Mal_{\m1}^-=\Mal_{\m1}^-\BT X=K[\m1]$. From the lemma ~\ref{oper-bxt} (iv) then we have that $\Mal_{\m1}\BT X^-=X^-\BT \Mal_{\m1}=K[\m1]^-=K[\m1]$. Hence $\Mal_{\m1}$ is also reversible.

(iii) From lemma~\ref{crit-P} and definition of the response matrix $\Hal$ for it (see lemma~\ref{gen}) we have $\Mal_g=\Mal_{\m1}\BT \Hal_g,g\in\m g$. Then $\Mal'_g=(A\BT\Mal_{\m1})\BT\Hal_g$.

Since $A$ is reversible the system of rows $\{A\BT\Mal_g|g\in\m g\}$ of the matrix $\Mal'$ has the same rank as the system of rows of the matrix $\Mal$. Therefore $\Mal'$ is non-singular matrix.

Finally,  from lemma~\ref{crit-P} we get that $\Mal'$ is a universal matrix with the response matrix $\Hal$.
\bx

\begin{cor}
Let $\Mal$ be a universal matrices representing an automorphism $\Theta$ of $\m M$ generated by an automorphism $\theta$ of the index group $\m g$. Then $\Mal_g^f=\Mal_{\m1}^{f\,\ovl{\theta^{-1}(g)}}$. The rows of the response matrix for $\Mal$ constitute the set $\{K[g]|g\in\m g\}$.
\end{cor}

\pr First consider the response matrix $\Hal$ for $\Mal$. As we know $\Hal_{\m1}=K[\m1]$. Also since $\Cal(\Theta(T))=\Mal^{-1}\Cal(T)\Mal$, for $T=K[g]$ we obtain
$\Mal\Cal(H[g])=\Cal(K[g])\Mal$ and deduce further that $H[g]=\Theta(K[g])=K[\theta^{-1}(g)],g\in\m g$.
 On the other hand $\Mal_g=\Mal_{\m1}\BT H[g]$. Therefore $\Mal_g^f=(\Mal_{\m1}\BT K[\theta^{-1}(g)])^f= \Mal_{\m1}^{f\,\ovl{\theta^{-1}(g)}}$.
\bx

{\sl Note:} if $\theta $ is a functional index-permutation then the corresponding automorphism of the index group is $\ovl{\theta(\m1)}\theta(\cdot)$. Therefore in terms of the functional index-permutations for the rows of the universal matrix  representing $\theta$ we have  $\Mal_g^f=(\Mal_{\m1}\BT K[\theta^{-1}(g)])^f= \Mal_{\m1}^{f\,\ovl{\theta^{-1}(\theta(\m1)g)}}$.\bx
\footnote{Check this on examples.} \\

\begin{ex}{\rm
There exist non-trivial response matrices proving that in general $\m i(\m M)\neq\m l(\m M)$.
We call response matrix {\sl full} if the only its row having a form $K[g]$ is the first one.
Table~\ref{A15} shows examples of full response matrix for $q=7,15$.
}\bx \end{ex}
{\tiny
\begin{table}\label{A15}
\caption{Full response matrices for $q=7$ (left) and $q=15$ (right).}
$\left(\begin {array}{ccccccc}\label{A15}
1&0&0&0&0&0&0\\
\noalign{\medskip}0&1&0&1&0&1&0\\
\noalign{\medskip}0&0&1&1&0&0&1\\
\noalign{\medskip}0&1&1&1&0&1&1\\
\noalign{\medskip}0&0&0&0&1&1&1\\
\noalign{\medskip}0&1&0&1&1&1&1\\
\noalign{\medskip}0&0&1&1&1&1&1
\end {array}\right),
\left(\begin {array}{ccccccccccccccc} 1&0&0&0&0&0&0&0&0&0&0&0&0&0&0\\
\noalign{\medskip}0&1&0&0&0&1&1&0&0&1&0&1&0&1&1\\
\noalign{\medskip}0
&0&1&1&0&0&0&1&0&0&1&1&1&1&0\\
\noalign{\medskip}0&1&1&0&1&0&1&1&1&0&0&0
&0&1&0\\
\noalign{\medskip}0&0&0&0&1&1&1&1&0&1&0&1&0&0&1
\\\noalign{\medskip}0&1&1&0&1&0&0&1&1&0&1&0&0&1&0\\
\noalign{\medskip}0
&1&1&0&1&0&0&0&1&0&0&1&1&0&1\\
\noalign{\medskip}0&0&1&1&0&1&1&0&1&1&1&0
&1&1&0\\
\noalign{\medskip}0&0&0&1&0&0&0&1&1&0&1&0&1&1&1
\\
\noalign{\medskip}0&1&1&1&1&0&0&0&1&0&0&1&0&0&1\\
\noalign{\medskip}0
&1&1&0&1&1&0&0&1&0&0&1&0&0&1\\
\noalign{\medskip}0&1&0&1&1&1&1&0&0&1&1&0
&1&0&1\\
\noalign{\medskip}0&1&1&0&1&0&0&1&1&1&0&0&0&1&0
\\\noalign{\medskip}0&0&1&1&0&1&1&1&1&1&1&0&1&0&0\\
\noalign{\medskip}0
&1&0&1&1&1&1&0&0&1&1&1&1&0&0
\end {array} \right)
$
\end{table}
}

\subsection{Regular isomorphisms}

Let $\aut{\m M}$ be the complete set of automorphisms of a monoid $\m M$, whereas $\m l(\m M),\m i(\m M)$ are respectively sets of linear automorphisms and automorphisms defined by automorphisms of  the index group (we call them {\sl index automorphisms}).

Given class $C\subseteq\aut{\m M}$, we denote as $\m I[C]$ the set of pairs ${T,H}$ of elements of $\m M$ such that there exists $\varphi\in C$ which translate $T$ into $H$. This means $\STDm T\approx\STDm H$ and therefore we can consider $\m I[C]$ as a class of isomorphisms of automata revealed by automorphisms from $C$.

There exist one obvious set of isomorphisms on HLCA being based on reversibility of rules in $\m M$.

\begin{theo}
If $T\in\m G$ then $\STDm T\approx\STDm{T^{-1}}$.
\end{theo}
\pr Indeed, $v\BT w=K[\m1]\iff w^-\BT v^-=K[\m1]$, i.e. $(v^-)^{-1}=(v^{-1})^-$. From here $(v,w)\in\STDm T\iff w=v\BT T^-\iff  w\BT(T^-)^{-1}=v\iff w\BT(T^{-1})^-=v\iff (w,v)\in\STDm{T^{-1}}$.

On the other hand if $T$ is reversible then it acts on set $\m V$ of states as 1-1-mapping and therefore its diagram is a graph consisting of cycles that are oriented in a way. If we revert the orientation we get an isomorphic graph which (as we saw just above) will be the diagram of the rule $T^{-1}$. \bx

The set of pairs $\m I_{\m G}=\{\{T,T^{-1}\}|T\in\m G,T\neq T^{-1}\}$ can extend isomorphisms generated by $\m l(\m M)$ as the example~\ref{q=7} shows. Therefore for $C\subseteq\aut{\m M}$ we denote $\m I[C]^+$ the set of isomorphisms of ACA on $\m g$ which is the closure of $\m  I[C] \cup\m I_{\m G}$.

Despite the mapping $T\mapsto T^{-1}$ for a concrete $T\in\m G$ could not be (generally speaking) extendable up either to an automorphism of $\m M$\footnote{It would be good to provide an example proving this.} or even to automorphism of $\m G$, the reflection $\varrho:T\mapsto T^{-1}$ was shown in lemma~\ref{refl} as the isomorphism between $\m G$ and the co-group $\m G^{\circ}$. Thus the extension $\aut{\m M}^+$ of $\aut{\m M}$ also relates to the isomorphisms of underlying algebraic structures.

\begin{cor}\label{inclusion}
$\m i(\m M)\subseteq\m l(\m M)\subseteq\aut{\m M}$.
\end{cor}

\pr  See theorem~\ref{ind-lin}. \bx

As examples show in general even for cyclic groups all inclusions in the corollary~\ref{inclusion} are proper.
\begin{ex}\label{q=7}
{\rm
If $\m g$ is a cyclic group of order 7 there are 12 different response matrices which do not represent index-permutations.
Table~\ref{A15} above shows one (left part). This matrix for pairs  $(1111000, 1001000)$, $(0101111,0100110)$ translates the left rule  into the right and therefore rules in each pair have isomorphic diagrams. However this could not be shown by index permutations because the number of units for vectors in each pair are different. Also it appears that the number of isomorphic classes in this case is 12 whereas there are 28 classes of isomorphism induced by index automorphisms and only 20 classes of isomorphism induced by linear automorphisms of $\m M$.This means that
\beas
\m i(\m M)\subsetneq\m l(\m M)\subsetneq\aut{\m M}
\eeas
for this case.  Moreover after the closure of the set of isomorphisms generated by linear automorphisms with isomorphisms generated by inversions of elements of $\m G$\footnote{Actually decreasing of the number of classes is enforced by isomorphisms of  rule with number $67$ to the inverse rule whose number is $118$ and $50$ to the inverse rule $87$.} we arrive at 18 classes of isomorphisms. Thus this example shows that in general the proper inclusions could hold:
\beas
\m I[\m i(\m M)]\subsetneq\m I[\m l(\m M)]\subsetneq\m I[\m l(\m M)]^{+}\subsetneq\m I[\aut{\m M}].
\eeas\bx}
\end{ex}

{\bf Definition:}{\sl\
We call any isomorphism of automata $\Acal(T)$ and $\Acal(L)$ {\sl regular} if it belongs to $\m I[\aut{\m M}]^+$. Also a group $\m g$ is called regular for $\m F_p$ if all automorphisms of automata on it are regular.
}

\begin{remar}{\rm It is important that any regular isomorphism is produced by an automorphism of the monoid and the isomorphism $\m G\to\m G^{\circ}$ and therefore is defined by an 1-1-mapping of one system of generators of $\m M,\m G$ onto another whereas in the definition of automata isomorphism we are talking about a much larger class of permutations of $\m V$.}\bx
\end{remar}

The next statement allows to show that some index groups are not regular.

\begin{theo}\label{not-reg}
Suppose for $\m g,\m F_p$ there exist two elements $v,w\in\m M\setminus\m G$ whose state transition diagrams are isomorphic but the numbers of solutions in $\m M$ to equations $x\BT x=v$ and $x\BT x=w$ are different. Then $\m g$ is not regular for $\m F_p$.
\end{theo}

\pr Suppose $|\{x|x\BT x=v\}|>|\{x|x\BT x=w\}|$.
Any automorphism $\varphi\in\aut{\m M}$ such that $\varphi(v)=w$ should translate the solutions to the equation $x\BT x=v$ into the solutions to $x\BT x=w$. This contradicts to the condition that $\varphi$ is 1-1-mapping on $\m V$. Thus $\{v,w\}\notin\m I[\aut{\m M}]$.

The state transition diagrams of reversible rules consist of cycles, whereas the diagrams of irreversible rules should have dangled vertices because of singularity of them as linear operators. Therefore there is no reversible rule with the diagram isomorphic to the diagram of an irreversible rule. This means (theorem~\ref{aut->iso}) that the classes $V,W$ of elements automorphic to $v,w$ correspondingly consist completely of irreversible rules.

On the other hand adding a pair $\{s,t\}\in\m I_{\m G}$ we can glue some two classes of automorphism of elements including $s$ and $t$ respectively. But these classes consist of reversible elements as elements $s,t$ are. Therefore the closure of $\m I[\aut{\m M}]$ with $\m I_{\m G}$ does not influence the classes of automorphism of irreversible elements.
\bx

\section{Automata isomorphisms for $\m F_2$ and small groups}

Results and examples from this section mostly relate to the case of the simplest field $\m F_2$, i.e. $p=2$. In this case we use to say simply "$\m g$ is (is not) regular".

\subsection{Case of cyclic $\m G$. Conjecture}

First it could be only if $\m g$ is cyclic: indeed, any subgroup of cyclic group is cyclic and there is an injection of $\m g$ into $\m G$.

Let $\m c_n$ be a cyclic group of order $n$. For the basic field $\m F_2$ we have then $\m1=0$ and $K[g], g\in\{0,\dots,n-1\}$ is a vector whose all components are equal 0 excepting $g$-th which is equal to 1. We also use $0,1$-words to write elements of $\m G$. So a word $0^{k_1}1^{k_2}\dots$ where $\sum_ik_i=n$ denotes a vector whose first $k_1$ components are zeroes, and these components are followed with $k_2$ components all equal to 1, etc.
Elements $\mathbf{0,1}$ play a special role further. By definition $\mathbf0=[0,\dots,0]=0^{n},\ \mathbf 1=[1,\dots,1]=1^n$.\footnote{We distinguish $\mathbf 1$ and $\m1$. As it was defined in the beginning the latter is the unit of the index group $\m c_n$.}

Let $T\in\m M$. We denote $L\in\m M$ as $T^{\star}$ iff $T^{\star}(i)=1-T(i), i=0,\dots,q-1$. Another way to write this is $T^{\star}=T+\mathbf 1$.

Also let $\pi(T)$ be a parity of $T\in\m V$, i.e. $\pi(T)=\sum_fT^f\in\m F_2$.

\begin{lem}\label{reg0} For any $T,L\in\m V$:\\
1) $\pi(T)\mathbf1=T\BT\mathbf1=\mathbf1\BT T$.\\
2) $\pi(T\BT L)=\pi(T)\pi(L)$.
\end{lem}

\pr
1) First of all, due lemma~\ref{oper-bxt} (vi) $T\BT\mathbf1=\mathbf1\BT T$.\footnote{For commutative index group this also follows from commutativity $\BT$.}
Now, $\pi(T)\mathbf1|^i=[(\sum_jT^j)\mathbf1]^i= \sum_jT^j\mathbf1^{i-j}=(T\BT\mathbf1)^i$.

2) $\pi(T\BT L)=\sum_j(T\BT L)^j=\sum_j\sum_iT^iL^{j-i}= \sum_iT^i[\sum_jL^{j-i}]=\sum_iT^i\pi(L)=\pi(L)\pi(T).$

\bx

\begin{lem}\label{reg1}
Let $\m g$ is of an odd order $n$. \\
1) $T\in\m M\implies \pi(T)\neq\pi(T^{\star})$.\\
2) $T\in\m G\implies\pi(T)=1$.\\
3) $\m G^{\star}=\{T^{\star}|T\in\m G\}$ is a subgroup of $\m M$ isomorphic to $\m G$.
\end{lem}

\pr
1) $\pi(T)=\sum_{i=0}^{n-1}T^i=\sum_{i=0}^{n-1}(1-(T^{\star})^i)=n- \pi(T^{\star})$. From here $\{\pi(T),\pi(T^{\star}\}=\{0,1\}$.

2) If $T$ is reversible then $\det[\Cal(T)]=1$.
However as it is known for circulants\cite{Ryzhik} $\det[\Cal(T)]$ is multiple to $\pi(T)$. Hence $\pi(T)\neq0$.

3) Since $T^{\star}=T+\mathbf1$ and lemma~\ref{oper-bxt} (iii)we can proceed as follows:
\beas
T^{\star}\BT L^{\star}=(T+\mathbf 1)\BT  (L+\mathbf1) =T\BT L+T\BT\mathbf1+L\BT\mathbf1 +\mathbf1\BT\mathbf1=\\
(T\BT L)^{\star}+[\pi(T)+\pi(L)]\mathbf1
\eeas
Therefore dealing with vectors $T,L$ of the same parity we can write
\beas
T^{\star}\BT L^{\star}=(T\BT L)^{\star}.
\eeas
This with the fact that $(\cdot)^{\star}$ is 1-1-mapping on $\m V$ yields that $(\cdot)^{\star}:\m G\to \m G^{\star}$ is an isomorphism.
\bx

In order for distinguishing between a degree $m$ of power $\underbrace{T\BT\dots\BT T}_{m\text{ times}}$ of a vector $T\in\m M$ and the $m$th component of the row-vector we denote $\underbrace{T\BT\dots\BT T}_{m\text{ terms}}$ as $T^{\BT m}$.

\begin{lem}\label{max-cyc}
Let positive integer $r$ is the minimal number such that $T^{\BT(r+1)}=T$ for a vector $T\in\m M$. Then $r$ is the length of the maximal cycle in $\STDm T$. In particular, if $T\in\m G$, i.e. is reversible, then the order of $T$ in $\m G$ is equal to the length of the maximal cycle in $\STDm T$.
\end{lem}

\pr
By the lemma~\ref{oper-bxt} (i),(iv) and theorem~\ref{lem2} we have
\bea\label{mcyc}
\begin{cases}
T^{\BT (m+1)}\BT A=T\BT A& \iff \\
T^{\BT m}\BT (T\BT A)=T\BT A& \iff \\
(A^-\BT T^-)\BT(T^-)^{\BT(m)}=
(A^-\BT T^-)& \iff \\
T^{\BT m}*(A^-\BT T^-)=(A^-\BT T^-)&
\end{cases}
\eea

Now, for any $B$ that belongs to a cycle in $\STDm T$ there exists an vector $A$ such that $B=T\BT A$. From here the length of any cycle in the diagram of the rule $T$ does not exceed $r$.

On the other hand, substituting $A:=K[\m1]$ in (\ref{mcyc}) and using (ii) from the lemma~\ref{oper-bxt} we arrive at the conclusion that the length of the cycle including $T^-$ in $\STDm T$ is not lesser $r$ because $r$ is the minimal number such that $T^{\BT (r+1)}=T$.
\bx

\begin{lem}\label{extent}
Let $H,V$ be sub-semigroups of the monoid $\m M$ without mutual elements and $\varphi: H\to H$ is an automorphism of $H$. If there exists an isomorphism $\gamma:V\to H$ such that $\gamma^{\sigma}(h\BT v)=h\BT\gamma^{\sigma}(v), h\in H,v\in V,\sigma\in\{-1,1\},$ then $\varphi$ can be extended to an automorphism of $H\cup V$.
\end{lem}

\pr
We define $\Phi:H\cup V\to H\cup V$ by
\bea
\Phi(z)=
\begin{cases}
\varphi(z),& z\in H,\\
\gamma^{-1}\varphi\gamma(z),& z\in V.
\end{cases}
\eea
To prove that $\Phi$ is an automorphism of $H\cup V$ first note that $\Phi$ is 1-1-mapping because the conditions that $H\cap V=\emptyset$ and $\varphi,\gamma$ are 1-1-mappings with $H$ being the set of their values. Now,
\beas
\Phi(h\BT v)=\gamma^{-1}\varphi\gamma(h\BT v)=_{\text{condition for $\gamma$ with $\sigma=1$}}\gamma^{-1}\varphi(r\BT\gamma(v))=\\ \gamma^{-1} (\varphi(r)\BT\varphi(\gamma(v)))= _{\text{condition  for $\gamma$ with $\sigma=-1$}}
\varphi(r)\BT\gamma^{-1}\varphi\gamma(v)=\Phi(r)\BT\Phi(v).
\eeas
\bx

\begin{theo}\label{reg-theo}
Assume that \\
(i) the number $n=|\m g|$ is odd;\\
(ii) $|\m G|=2^{n-1}-1$;\\
(iii) for each two elements $T,H\in\m G$ of an equal order there exists an automorphism $\varphi$ of $\m G$ such that $\varphi(T)=H$.\\
Then all isomorphisms of ACA on $\m g$ are regular.
\end{theo}

\pr
Since $T\in\m M$ with even parity cannot be reversible, the condition (ii) results that all elements of $\m M$ with odd parity (excluding $\mathbf1$ whose circulant has a determinant equal to 0) constitute $\m G$ and therefore are reversible.

All elements of $\m M$ of even parity excluding $\mathbf 0$ constitute a subgroup (more exactly: sub-monoid) $\m G^{\star}$ of the monoid and by lemma~\ref{reg1} $\m G\simeq\m G^{\star}$. The mapping $T\mapsto T+\mathbf 1$ serves as an isomorphism $\gamma$. Indeed, $T\mapsto T+\mathbf 1=T^{\star}$ and clearly the identity $\gamma(\gamma(T))=T$ holds. That is $\gamma=\gamma^{-1}$. In addition
\beas
T\BT(H+\mathbf 1)=T\BT H+T\BT\mathbf 1=T\BT H+\mathbf 1.
\eeas
This means $T\BT\gamma(H)=\gamma(T\BT H)$ and therefore by lemma~\ref{extent} any automorphism $\varphi$ of $\m G$ is extendable to an automorphism $\Phi$ of $\m G\cup\m G^{\star}$.

Then it obviously can be extended to an automorphism $\Phi'$ of $\m G\cup\m G^{\star}\cup\{\mathbf0\}$ setting $\Phi'(\mathbf 0)=\mathbf 0$ because $T\mathbf0=\mathbf0$.

Finally we define $\Phi''(X)=\Phi'(X)$ if $X\neq\mathbf1$ and
$\Phi''(\mathbf1)=\mathbf1$. Let show that $\Phi''\in\aut{\m M}$. For that it is enough to consider the action of $\Phi''$ of products of kind $T\BT\mathbf 1$. Since $\Phi''$ preserves parity $\pi(T)$ of element $T$ and by lemma\ref{reg0} $T\BT\mathbf1=\pi(T)\mathbf1$, the equality $\Phi''(T\BT\mathbf 1)=\Phi''(T)\BT\Phi''(\mathbf1)$ holds.
Hence $\Phi''\in\aut{\m M}$.

From lemma~\ref{max-cyc} it follows that elements $T,H$ of different orders cannot have isomorphic diagrams. This means that no automorphism of $\m M$ exists which translates $T$ into $H$.

On the other hand from (iii) we have that for any elements $T,H\in\m G$ of the same orders there exists an automorphism $\varphi:\m G\to\m G$ s.t. $\varphi(T)=H$. As we showed, this automorphism of $\m G$ can be extended to an automorphism of $\m M$.

Because of the isomorphism $\gamma:\m G^{\star}\to\m G$ the same is true for elements $T,H\in \m G^{\star}$.

In addition, $\{\mathbf0\},\{\mathbf1\}$ are singleton classes of isomorphism \cite{Bul2}.

Finally the extension of the set of isomorphisms $\m I[\aut{\m M}]$ by $\m I_{\m G}$ does not yield anything of new because both $T,T^{-1}, T\in\m G,$ have the same order and we can refer to the condition (iii).

Thus all isomorphisms of ACA on $\m g$ are regular.\bx

The next lemma just recalls a well known fact:
\begin{lem}\label{aut_cyc}
For any two elements of the same order in a finite cyclic group there exists an automorphism of the group translating one of these elements into another.
\end{lem}

\pr
Any cyclic group of order $n$ is isomorphic to the standard cyclic group $\Mal_n=\la\{0,\dots,n-1\},+_{\mm n}\ra$. So we can reason about this group. Now, an element $j$ of the group has order $k\iff \gcd(n,j)=\frac nk$, in other words $k=\frac{n}{\gcd(n,j)}$. Indeed, the order $k_j$ of $j$ is the least number $r$ such that $jr=sn$ for a number $s$. On the other hand $k_j|n$ because orders of elements divide the order of the group (Lagrange theorem ). From here $j=s\frac n{k_j}=st, t\in\mathbb Z^+$. Thus $t$ is the maximal divisor $j$ such that is simultaneously a divisor of $n$ ($t=\frac n{k_j}$). Hence $t=\gcd(n,j)$ and $k_j=\frac{n}{\gcd(n,j)}$.

Let $a,b$ have the same order, i.e. $\gcd(n,a)=\gcd(n,b)=d$. If so then numbers $\frac ad,\frac bd$ are both relatively prime with $n$ and therefore are generators for the group $\m M_n$. Let us define $\varphi:[i(\frac ad)\mm n]\mapsto [i(\frac bd)\mm n]$ where $i$ runs over $\{0,\dots,n-1\}$. This mapping is an automorphism translating $a$ into $b$ because $a=d(\frac ad)$ and $b=d(\frac bd)$.
Indeed, for any $x,y\in\{0,\dots,n-1\}$ there exist multipliers $X,Y$ such that $\frac adX=x,\frac adY=y$. Then $\varphi(x+y)=\varphi(\frac adX+\frac adY)=\varphi((X+Y)\frac ad)=(X+Y)\frac bd=X\frac bd+Y\frac bd=\varphi(x)+\varphi(y)$. Finally $\varphi$ is a bijection because both $\frac ad,\frac bd$ are relatively prime with $n$ that enforces both numbers $i(\frac ad)\mm n,i(\frac bd)\mm n$ run without repetitions the set $\{0,\dots,n-1\}$ when $i $ runs over this set. \bx

The next statement formulates conditions when elements of $\aut{\m M}$ are compositions of automorphisms of $\m G$ and an isomorphism between $\m G$ and $\m G^{\star}$.
\begin{cor}
If $\m G$ is a cyclic of order $2^{|\m g|-1}-1$, then any two different automata $\Acal(f),\Acal(g),f\neq g,$ on $\m g$ have isomorphic diagrams iff there exists an automorphism $\psi$ of $\m G$ such that $\psi(f)=g$ or $\psi(f^{\star})=g^{\star}$.
\end{cor}

\pr
If $\m g$ is not cyclic, $\m G$ is not cyclic as well (any subgroup of cyclic group is cyclic). And if 2 is not a primitive root modulo $|\m g|$ then $\m G$ cannot be a cyclic group of the order $2^{|\m g|-1}-1$ because all elements have orders lesser this number. Hence $|\m g|$ is an odd prime and we can apply theorem~\ref{reg-theo} getting that all isomorphisms are regular.

On the other hand, because the reverse
element to an element $g$ of a group has the same order, from the lemma~\ref{aut_cyc} with the method of extension of automorphisms of $\m G,\m G^{\star}$ that we applied in the proof  of theorem~\ref{reg-theo} on the basis of lemma~\ref{extent} we have that there is no proper extension of the class of automorphisms of $\m G$ by reversion of elements.
Now, as we saw above, $\m G,\m G^{\star}$ are isomorphic and any automorphism of $\m M$ translates $\m G$ into $\m G$ and $\m G^{\star}$ into $\m G^{\star}$. Thus any automorphism of $\m M$ can be constructed from automorphisms of $\m G$.
Finally, since $f\neq g$ we note that both rules $f,g$ should belong to the same set among $\m G,\m G^{\star}$.
\bx

\begin{ex}\label{prim-root}{\rm
In cases when index group $\m g$ is a cyclic group of order $q$ that 2 is a primitive root modulo $q$, computations yield that in each of the
cases $q\in\m R$ where $\m R=\{3,5,11,13,19,29,37,53,59,61,67,83,101,107,131\}$\footnote{This is the initial segment of the sequence A001122, see $\la$ http://www.research.att.com/~njas/sequences/A001122$\ra$
} there exists an element of the corresponding monoid over $\m F_2$ having order $2^{q-1}-1$. (For instance, the 01-vectors with numbers $1,21,11,13,19$ have orders $4,15,1023,4095,$ and $262143$ in monoids with $|\m g|=3,5,11,13,19$ respectively.) This means that in these cases the groups of reversible elements of the monoids for cyclic groups of the orders belonging to $\m R$ are also cyclic groups of order $2^{q-1}-1.$ So the lemma~\ref{aut_cyc} holds for the groups of reversible elements and the theorem~\ref{reg-theo} says that all isomorphisms of ACA on the cyclic groups of orders $q\in\m R$ are regular.
\bx}
\end{ex}

\begin{cor}
If $\m G$ is a cyclic, then $\m i(\m M)=\m l(\m M)$.
\end{cor}

\pr This is because in this case there exists the unique subgroup (consisting of elements of $\mathbf K=\{K[g]|g\in\m g\}$) of order $|\m g|$ in $\m G$. \bx

Thus for all index groups from Example~\ref{prim-root} all linear automorphisms are index-permutations.

\noindent{\bf  Conjecture. }
{\sl If $\m g$ is cyclic group $\m c_q$ of an order $q$ such that 2 is a primitive root modulo q, then $\m G$ is a cyclic group of order $2^{q-1}-1$ and therefore all isomorphisms of ACA on $\m g$ are regular.}

\begin{remar}
{\rm
According to the known Artin conjecture on the set of primes (see the references in \cite{Artin-wiki}) there exist infinitely many primitive roots for any prime $p$. This Artin conjecture follows from Generalized Riemann hypothesis \cite{Hooley}.}\bx
\end{remar}

\subsection{Cases of non-cyclic $\m G$}
There are index groups $\m g$ for which the hyper group $\m G$ is not cyclic but nevertheless all isomorphisms of ACA on them are regular. These are for example $\m D_2,\m c_6,\m c_7$.
On the other hand not all isomorphisms of ACA are regular for even simple index groups $\m g$ like $c_4,\m D_3$, etc.

\begin{ex}{\rm There are only 2 groups (within isomorphism) of order 4.\\
{\sl Case 1: $\m g=\m c_4$}. In this case $\m G$ is the group $\m c_4\times\m c_2$ that is one of three commutative groups of order 8. All rules are partitioned into 8 classes of true isomorphisms of their diagrams. To check if automorphisms of $\m M$ reveal $\m I[\m M]$ it is possible to check all systems of 3 generators $\alpha,\beta$ for $\m G$ of orders 2 and 4 respectively and one $\gamma$ from $\m M\setminus\m G$. By "brute force" it is not difficult to check all candidates to be automorphisms of $\m M$ (there are lesser 10,000) of them and find $\m I[\aut{\m M}]$. It appears that the isomorphism of automata $\Acal(5)$ and $\Acal(10)$ is not revealed by $\aut{\m M}$ as well as the isomorphism of automata $\Acal(2)$ and $\Acal(13)$. Therefore $\m J[\aut{\m M}]$ produces 4 classes  of isomorphisms
$\{5\},\{10\},\{2\},\{13\}$ instead of true two $\{5,10\},\{2,13\}$. All other classes of isomorphic rules are equal to the true classes of isomorphic rules. Addition of $\m I_{\m G}$ does not yield anything of new because rules 5,10 are not reversible and 2 and 13 have the same order 2, that is $(0010)^{-1}=0010,\ (1110)^{-1}=1110$, and there is no pair in $\m I_{\m G}$ including any of these elements (pay attention that all other true classes of isomorphic rules are revealed by $\aut{\m M}$).
The absence of an automorphism $\varphi\in\aut{\m M}$ translating rule 2 into rule 13 is confirmed (see theorem~\ref{not-reg}) by the fact that the equation $X\BT X=1110$ has no solution in $\m M$ whereas there are 4 solutions for $X\BT X=0010$.
From here it follows that the group $\m c_4$ {\sl is not regular}.\\
{\sl Case 2: $\m g$ is the "Klein four group"}. This time $\m G$ is isomorphic to $\m c_2^3$; there are 6 true classes of isomorphic rules. We used generating systems of 6 elements for $\m M$ and found a few isomorphisms from $\aut{\m M}$ providing the true isomorphic classes of rules. So here $\m I[\aut{\m M}]=\m I[\m M]$ and the group is regular.
}\bx
\end{ex}

\begin{ex}{\rm
For {\sl non-abelian group of order 6 ($\m D_3$)} we have isomorphic rules $T,H$ with numbers 39 and 52 respectively. However there could not exist (see theorem~\ref{not-reg}) any $\varphi\in\aut{\m M}$ such that $\varphi(T)=H$ because the equation $X\BT X=T$ has 2 solutions, whereas $X\BT X=H$ has 8 solutions.

In contrast to this the abelian group of order 6 (it is  $\m c_6$) is   regular.
}\bx
\end{ex}

\begin{ex}{\rm
For the case of the {\sl group $\m Q$ of quaternions} as $\m g$ let $T,H$ are irreversible rules with numbers $9$ and $144$. They are isomorphic rules. However the equations $X\BT X=T$ and $X\BT X=H$ have 16 and 48 solutions respectively. The theorem~\ref{not-reg} is applicable in this case as well and we conclude that $\m Q$ is not a regular group.}\bx
\end{ex}

\begin{table}[here]
\caption{Index-groups $\m g$ are cyclic.
}
\begin{tabular}{|l|r|r|r|r|r|r|c|}
\hline
$\m g$&$\m c_1$&$\m c_2$&$\m c_4$&$\m c_6$&$\m c_7$&$\m c_8$&$\m c_q,q\in\m R$\\
\hline
Regular?&yes&yes&no&yes&yes&no&yes\\
\hline
\end{tabular}
\end{table}

\begin{table}[here]
\caption{Index-groups $\m g$ are non-cyclic.
}
\begin{tabular}{|l|r|r|r|r|r|}
\hline
$\m g$&$\m D_2$&$\m D_3$&$\m c_4\times\m c_2$&$\m D_4$&$\m Q$\\
\hline
Order&4&6&8&8&8\\
\hline
Commutative?&yes&no&yes&no&no\\
\hline
Regular?&yes&no&no&no&no\\
\hline
\end{tabular}
\end{table}

The open question is whether
at least some non-regular isomorphisms have a combinatorial nature not reducible to the symmetries of the underlying algebraic structures $\m g, \m G,\m M$.\\


\begin{thebibliography}{99}

\bibitem{Ulam-Neumann} J. von Neumann, A.W. Burks. \textsl{Theory of self-reproducing automata}. Univ. of Illinois Press, 1966.

\bibitem{Wolf} S. Wolfram. \textsl{A new kind of science}. Wolfram Media, 2002.

\bibitem{linCA} P.P. Chaudhuri, D.R. Chowdhuri, S. Nandi, S. Chattopadhyay. \textsl{Additive Cellular Automata: theory and applications}. V.1, Willey-IEEE Computer Society Press, 1997.

\bibitem{Coxeter} H.S.M. Coxeter, W.O.J. Moser. \textsl{Generators and relations for discrete groups}. 3rd ed., Springer Verlag, N.-Y. 1972.

\bibitem{Weis} E.W. Weisstein. \textsl{Group convolution}. A Wolfram Web resource.\\  http://mathworld.wolfram.com/GroupConvolution.html

\bibitem{Clifford} A.H. Clifford and G.B. Preston. \textsl{The algebraic theory of semigroups}. Vol.1, AMS, 1961.

\bibitem{Rosen} K.H. Rosen. \textsl{Elementary number theory and its applications}. 3rd ed. Addison-Wesley, 1993.

\bibitem{Ryzhik} I.S. Gradstein, I.M. Ryzhik. \textsl{Table of integrals, series, and products}. 6th ed. Academic Press, 2000.


\bibitem{Artin-wiki} http://en.wikipedia.org/wiki/Artin${}_{-}$conjecture

\bibitem{Hooley} C. Hooley. \textsl{On Artin's conjecture}. J.Reine Angew. Math. {\bf 225}, 1967, 209-220.

\bibitem{Bul1} V.K. Bulitko, B. Voorhees, V.V. Bulitko. \textsl{
Discrere Baker Transformations for Linear Cellular Automata Analysis}. Journal of Cellular Automata, {\bf1}, Num. 1, 2006, 40-70.

\bibitem{Bul2} V. Bulitko, B. Voorhees. \textsl{Index permutations and classes of additive CA rules with isomorphic STD}, Journal of Cellular Automata, (in print).


\end{thebibliography}
\end{document}